# Experimental demonstration of ions induced electric field in perovskite solar cells

Zeguo Tang[1,*] & Takashi Minemoto [2]

[1]Ritsumeikan Global Innovation Research Organization, Ritsumeikan University, 1-1-1 Nojihigashi, Kusatsu, Shiga 525-8577, Japan

[2]Department of Electrical and Electronic Engineering, Ritsumeikan University, 1-1-1 Nojihigashi, Kusatsu, Shiga 525-8577, Japan

Email: tangzg@fc.ritsumei.ac.jp

**Abstract**: The anomalous hysteresis is reported in photo current density-voltage (*J-V*) curves of perovskite solar cells. The origins responsible for hysteresis are primarily attributed to ferroelectricity, trapping/de-trapping, and ions migration. Meanwhile, a switchable photovoltaic effect is disclosed in lateral structure perovskite solar cells, where a p-i-n structure is formed after poling the device with a reverse bias. Considering the normal sandwiched structure of perovskite solar cells, i.e. the intrinsic perovskite layer is situated in an electric field built by p-type (Spiro-OMeTAD) and n-type ($TiO_2$) layers, the poling process also works in such devices. The migration of ions will induce a new electric field with opposite direction as inherent built-in electric field. As a consequence, less collection of photogenerated carriers is predicted due to the built-in electric field is partially screened. Here, the ions induced electric field is experimentally demonstrated by exploring the external quantum efficiency (EQE) spectra under various forward voltage biases. The anomalous hysteresis observed on *J-V* curves is explained based on the insight of ions migrations under bias voltage. The external applied voltage presents significant influence on the stability of solar cells. Meanwhile, a novel phenomenon of negative capacitance is initially detected at low frequency, which is ascribed to the position reversal of negatively and positively charged ions under forward bias.

Organometal tri-halide perovskite solar cells draw considerable attraction due to the amazing advancement in last six years since the first cell with conversion efficiency of 3.8% was reported by Prof. Miyasaka[1]. The champion conversion efficiency has been improved to 20.1%[2], a value comparable to traditional compound thin-film solar cells of CuInGaSe$_2$ with champion efficiency of 21.7%[3] and CdTe with top efficiency of 21.5%[4]. At present, the most challenge for such cells is the instability, which will inhibit the cells to enter the market[5-7]. Meanwhile, the novel phenomenon termed hysteresis, namely, the conversion efficiency difference dependent upon the scan direction, scan rate as well as precondition prior measurement, is observed during measuring the *J-V* characteristics[8-20]. There are many literatures discussed the origins of hysteresis phenomenon. Michaele Grätzel et al. ascribed the reason of hysteresis to the charge carrier collection while Henry J. Snaith and M. D. McGehee et al. deemed the ions migration is responsible for the anomalous hysteresis[9-11]. Aron Walsh et al. accounted for microscopic polarization domains made considerable contribution to the anomalous hysteresis[12]. Jinsong Huang et al. attributed it to the trap defect at surface and grain boundaries of perovskite material[13]. Almost all of literatures were focused on the photo *J-V* characteristic and discussed the origins of anomalous hysteresis based on these results. Some paper even considered that the hysteresis is originated from photo-induced ions migration[11]. Despite of all these proposals, there is still a debate regarding the main origin of hysteresis.

Furthermore, an intriguing switchable photovoltaic effect is reported by Jingsong Huang[21, 22], where a p-i-n structure is formed after poling the lateral structure device with a reverse bias. Taking into account the structure of normal perovskite solar cells, namely, the intrinsic perovskite layer is sandwiched by a p-type (Spiro-OMeTAD) and n-type (TiO$_2$) layers, the perovskite layer is situated in a built-in electric field. Thus, the perovskite layer is undergoing a poling process and the migration of ions will occur in perovskite solar cells. To date, the mobile ions species are still debate, Jingsong Huang et al. observed the migration of MA$^+$ by photothermal induced resonance (PTIR) microscopy[22]; Simultaneously, Christopher Eames et al. suggested that I$^-$ was the most possible ions drifting in the built-in electric field by simulation[23]. Anyway, an ions induced electric field is supposed. In this contribution, the ions induced electric field is experimentally demonstrated via investigating the EQE spectra under forward bias. The hysteresis observed in photo and dark *J-V* curves is explained taking into account the alteration of total electric field in solar cells when voltage bias is applied on electrodes. Conversion efficiency degradation under different voltage biases reveals that the stability of such solar cells is significantly influenced by the electric field in device. Meanwhile, a novel phenomenon of negative capacitance is primarily detected in capacitance-voltage (*C-V*) characteristic, which is ascribed to reversal of ions under electric field.

## Results

**Ions induced electric field hypothesis and demonstration.** Perovskite solar cells with mesoporous structure of FTO/blocking $TiO_2$/mesoporous $TIO_2$/perovskite/Spiro-OMeTAD/Au are prepared for investigation, and the detailed fabrication processes are described in the experimental section. First, we illustrate the schematic of band diagram for perovskite solar cells with p-i-n structure in Fig. 1. In this manuscript, we assume both positively and negatively charged ions (cations and anions) migrate under electric field. Meanwhile, $MA^+$ represents all cations while $I^-$ depicts all anions. The intrinsic perovskite layer is sandwiched by a p-type Spiro-OMeTAD layer and an n-type $TiO_2$ layer. In other words, the perovskite layer is situated in a built-in electric field. As a result, $MA^+$ will shift towards $TiO_2$ side while $I^-$ will shift towards Spiro-OMeTAD side. Thus questions arise: Is there an ions induced electric field? Does it affect the inherent built-in electric field in the solar cells? The predictions of a reduced built-in electric field due to ionic screening and resultant less carriers collection efficiency have be made in many literatures[23, 24]. However, the direct demonstration of this effect has never been reported.

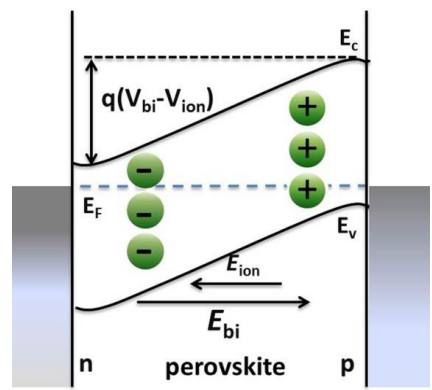

Fig.1 Schematic of band diagram for p-i-n structured perovskite solar cells at short circuit. + denotes cations while - presents anions.

As we know that the EQE spectrum measurement is an effective method to detect the carrier collection efficiency in solar cells. Herein, we conduct the measurement of EQE spectrum under different voltage biases. Figure 2a shows the relevant EQE spectra under forward bias for a perovskite solar cell with $(FAPbI_3)_{0.85}(MAPbI_3)_{0.15}$ as absorber layer. The corresponding *J-V* curve is shown in Supplementary Fig. 1. Intriguingly, a novel phenomenon is observed, i.e. the EQE spectra intensity is enhanced with the applied forward bias and reaches summit at bias of 0.5 V, as shown in Fig. 2a. For a clear view, Fig. 2b shows the comparison of EQE and related integrated short circuit current density ($J_{sc}$) under the bias of 0 and 0.5 V. Further increase of the bias results in the reduction of EQE intensity. However, the EQE intensity rises up again at long wavelength region when forward bias is larger than the inherent electric field. In principle, the EQE will decrease while a forward bias is applied on solar cells due to the reduction of total electric field. Nevertheless, here, the EQE enhancement under small forward bias is observed. In addition, the EQE spectra under reverse bias are also detected and the corresponding results are depicted in Supplementary Fig. 2. Evidently, the EQE spectra are enhanced with increasing reverse bias, which is ascribed to the increase of total electric field in device.

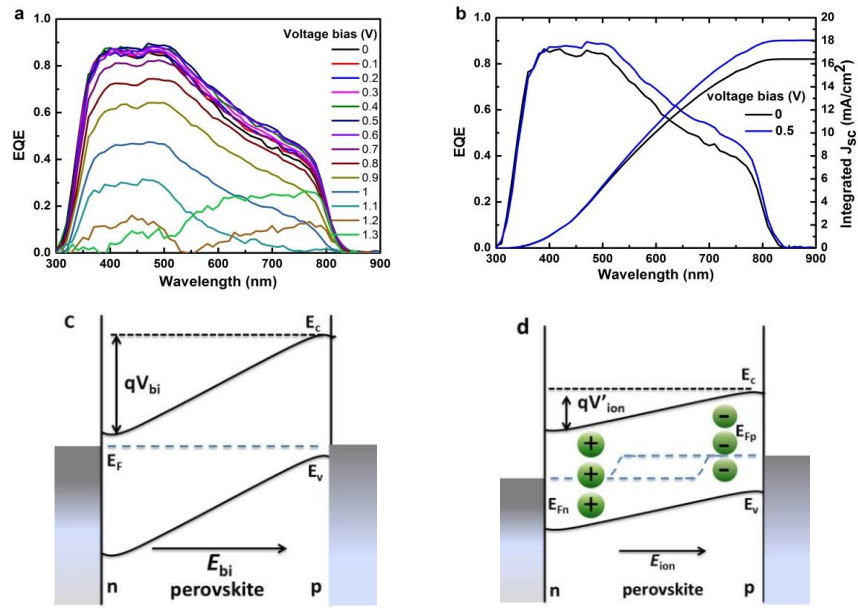

Fig.2 EQE spectra under different (a) forward voltage biases; (b) 0 V and 0.5 V biases and related integrated Jsc. Band diagram in perovskite solar cells under forward bias (c) equals to ions induced voltage; (d) more than inherent field.

To clearly explain this phenomenon, schematics of band diagram for p-i-n perovskite solar cells under forward bias are drawn in Figs. 2c and 2d. As shown in Fig. 1, an ions induced electric field is existed due to the migration of $MA^+$ and $I^-$ driven by the built-in potential. When a forward bias is applied, the electric field in device is altered. The external forward bias leads to ions return back to their initial locations, and then the ion-induced field is decreased. Hence, the electric field in the solar cell is enhanced and an intriguing consequence of enhanced EQE is observed. When the applied voltage equals to the ions induced electric field, the EQE intensity reaches maximum, suggesting the significant enhancement of carrier collection efficiency. It is worth noting that the integrated $J_{sc}$ calculated from EQE spectrum at 0 bias is smaller than that obtained from *J-V* curve. One of possible reason is that the built-in potential is partially screened by ions in perovskite layer and consequent reduction of total electric field in solar cells. As shown in Fig. 2b, the integrated $J_{sc}$ from the EQE spectrum under 0.5 V is comparable to the value obtained from *J-V* curves shown in Supplementary Fig.1. This finding directly supplies an evidence for the ions screening effect in perovskite solar cells[23, 24]. Next, the EQE intensity decreases when further increasing bias to over the ions induced electric field, which is attributed to the reduction of total electric field. However, the EQE intensity is enhanced again at long wavelength region while the forward bias is more than the inherent electric field. This is because the forward bias not only counteracts the inherent electric field but also drives the ions migration to form a new ion electric field with opposite direction, as shown in Fig. 2d. Photo generated carriers are extracted under this new ions electric field.

The p-i-n structure of perovskite solar cell determines that intrinsic perovskite layer is located in a field built by the p- and n-type layers. This is similar with the poling effect, reported by Jinsong Huang[21, 22], where a voltage is applied on the perovskite layer. After several durations, the self-doping is realized and p-i-n structure is formed in perovskite layer. *J-V* curve for such cell suggests that $V_{oc}$ is around 0.6 V, which is smaller than that of cells with normal sandwiched structure. For normal p-i-n sandwiched structure solar cells, the built-in potential is mainly determined by the p-type and n-type materials. However, the self-doping structured cell, the built-in potential is determined by degree of poling. The $V_{oc}$ obtained from poling treated cell is about 0.6 V, a value comparable with the forward bias that resulted in largest EQE in this study. Thus, an assertion can be made that the value of ion-induced electric field is dependent on the degree of poling. To certify this, EQE spectra of other two solar cells are evaluated and the corresponding results are shown in Supplementary Fig. 3a and 3b. The EQE spectra arrive at summit at the forward bias of 0.6 V for solar cell shown in Supplementary Fig. 3a, where the solar cell is based on $CH_3NH_3PbI_3$ material. In Supplementary Fig. 3b, the bias value caused the maximum EQE spectrum is 0.4 V. Furthermore, the EQE intensity in long wavelength does not present an expected enhancement. Considering the thicker mesoporous layer in this cell compared to that shown in Fig. 2a, the reason is possibly related to the thickness of perovskite layer.

To understand the influence of perovskite layer thickness on hysteresis clearly, the cross-sectional SEM images for cells shown in Fig.2a and Supplementary Fig. 3b are characterized and corresponding results are depicted in Supplementary Fig. 4a and b, respectively. As reported in many literatures, the mesoporous structured device involves two perovskite layers: a perovskite infiltrated mesoporous layer and a capped perovskite layer[14, 25]. Here, we can clearly find that the perovskite layer in Fig. 4a is thicker than that in Supplementary Fig. 4b. Taking into account the thinner mesoporous layer in Supplementary Fig. 4a than that in Supplementary Fig. 4b, the capped perovskite layer in Supplementary Fig. 4a is thicker than that in Supplementary Fig. 4b. From above results, a consequence can be drawn that the infiltrated perovskite layer is difficult to be induced by the electric field, hysteresis is mainly originated from the capped perovskite layer. The thicker of the capped perovskite layer, the severer of hysteresis. As evidence, the photo *J-V* curve for perovskite solar cell with thin perovskite layer is also measured, as shown in Supplementary Fig. 5. The conversion efficiency deviation between reverse and forward scan direction is slighter than that shown in Supplementary Fig. 1. Thus, we can explain the reason why the EQE spectra are enhanced just in long wavelength region when the large forward bias is applied on the solar cell. The short wavelength light is absorbed by $TiO_2$ and infiltrated perovskite layer while the long wavelength light is absorbed by the capped perovskite layer. The ions induced electric field is mainly formed in capped perovskite layer. If all of the perovskite materials are infiltrated in mesoporous $TiO_2$ layer, the perovskite materials are difficult to be induced. Hence, the EQE intensity in long wavelength is not raised up at larger forward bias. Additionally, it can also explain the reason why the hysteresis is serious in planer structured cells than in mesoporous structured ones[14, 20, 25].

**Hysteresis phenomenon in perovskite solar cell.**

Next, let us review the hysteresis observed in *J-V* curves. Up to now, almost all of papers reported anomalous hysteresis in photo *J-V* curves. However, from above results, we find that the ions migration occurs as soon as voltage bias is applied on the electrodes regardless of light condition. To verify the variation, *J-V* curves at dark are recorded under reverse scan direction (from open circuit voltage to short circuit current) and to take into account the distinct of mobility between ions and electrons, the delay time is varied to investigate the impact of ions migration on *J-V* curves. Figure 3a shows the corresponding result, it is clear that a crook, i.e. current density first increases and then decreases, is observed at the beginning of bias applied on electrodes.

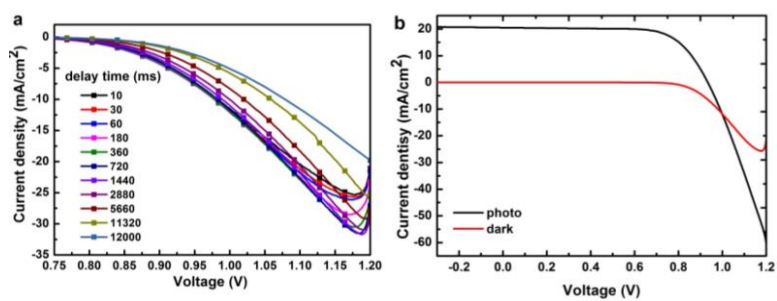

Fig.3 (a) *J-V* curves at dark condition recorded at different delay time. (b) Comparison of photo and dark J-V curves.

The $MA^+$ and $I^-$ will return back to their initial locations as soon as the bias applied on the device. Meanwhile, the electrons and holes will transport under the electric field. The ions have the same transport direction with the electrons and holes. When delay time is short, ions migration can't follow the alteration of electric field. A consequent crook is detected on dark *J-V* curve at the beginning of bias applied. The crook disappears when the delay time is extended to as longer as 12 s, a sufficient time for ions movement. The result suggests that ions movement is independent on light condition and the time duration of ions movement is about 12 s, which is comparable with other reports[11, 16].

Now, let us consider how the light illumination affects the hysteresis. There are some literatures reported that halide ions presented considerable mobilities in metal halides and metal halide perovskites[26-28]. In addition, photo illumination will accelerate the ions movement[29]. Moreover, it is well known that the photo-generated carriers will lead to the separation of quasi-Fermi level of hole and electron, an electric field with opposite direction with the built-in field will create[30]. This electric field will influence the migration of ions and then counteracts the ions induced electric field. Figure 3b shows *J-V* characteristics at dark and under illumination recorded under reverse scan direction at delay time of 30 ms and scan rate of 0.1 V/s. Obviously, a crook exists in the dark curve but disappears in the illumination case. We think two reasons are responsible for this result: one is the ions migration acceleration caused by photo illumination; another is the ions return back to their initial location under photo-generated field, resulting in the decrease of ions induced field.

As did in many literatures, *J-V* curves under illumination are recorded from –0.3 to 1.2 V with a scan rate of 0.1 V/s and a delay time of 30 ms. Figure 4a shows the corresponding curves and it is clear that a small deviation is observed on *J-V* curves obtained from the reverse and forward (from short circuit density to open circuit voltage) scan directions. Meanwhile, power conversion efficiency obtained from reverse scan is larger than that from forward scan, which is consistent with other reports[31-35]. To investigate the impact of scan rate on *J-V* curve, scan step is varied from 0.001 to 0.01 V to adjust scan rate. Fig.4b shows the *J-V* curves recorded under scan rate of 0.033, 0.1, 0.165 and 0.33 V/s, respectively. Clearly, hysteresis becomes serious with the increasing scan rate. It needs to point out that the curve obtained at scan rate of 0.1 V/s displays better performance than others, which is because this curve is recorded first. There are many papers reported that hysteresis will become severe over time[5, 9, 20]. To rectify the aging of perovskite solar cell on hysteresis, repeated scan is conducted. Figure 4c indicates that hysteresis actually become serious over scan times, which is attributed to the perovskite layer aging during the measurement. Combination of Fig. 1, the forward bias cancels out the ions induced electric field while the reverse bias strengthens it. When the voltage is swept from forward to reverse, the ions induced electric field is canceled out. The total electric field in solar cells is enhanced and then results in the increase of current density. However, the total electric field will reduce when the voltage is swept from reverse to forward. This is why the conversion efficiency got from reverse scan direction is larger than that obtained from forward scan direction. Moreover, the perovskite material is easily decomposed to $PbI_2$ and MAI. Light illumination speeds up the decomposition, and then leads to more ions[36]. This can explain the more severe hysteresis for the aged solar cells. Briefly, more and more evidence reveal that hysteresis is originated from ions migration in perovskite layer[10, 11, 35]. Considering the fact that the perovskite materials are mixed ionic-electronic conductors, hysteresis occurs due to the different mobility of electrons/holes and ions.

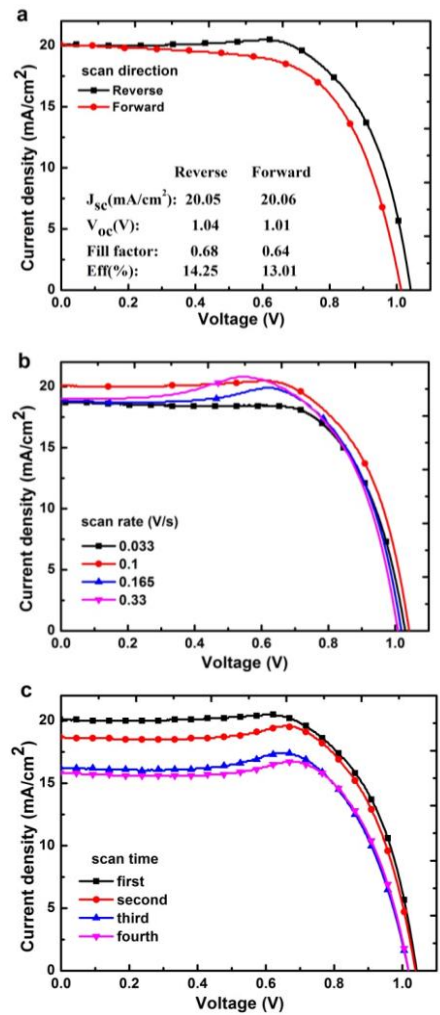

Fig.4 (a) *J-V* curves under forward and reverse scan direction. (b) Impact of scan rate on hysteresis. (c) *J-V* curves under repeated scan.

In addition, it is reported that the precondition of light illumination exhibits considerable influence on hysteresis. Here, to evaluate the influence of pre-illumination on ions migration, photo *J-V* curves are recorded with reverse scan direction from various starting voltage. Figure 5a shows the corresponding results with pre-illumination of more than 30 s. From this, we can find that the current density decreases at the beginning of voltage applied. As a comparison, the *J-V* curves without pre-illumination are also shown in Fig. 5b. The opposite trend is observed, i.e. the current increase at the beginning of voltage applied. In the case of pre-illumination, the photo-generated electric field is established due to the separation of quasi-fermi levels of holes ($E_{Fh}$) and electrons ($E_{Fn}$)[30]. Owing to the opposite direction of this field with the built-in electric field, the $MA^+$ and $I^-$ ions return back to their initial locations. In other words, the ions induced electric field is eliminated. The $MA^+$ and $I^-$ ions shift towards opposite direction with the holes and electrons do as soon as the bias is applied on the device. Therefore, a part of current is counteracted, and then the current density decreases. However, in the case without pre-illumination, the carriers are generated when light irradiates on device. If merely electrons make contribution to the current density, the curve should jump from dark current to photo current. However, the ions also drift under the electric field. The slower mobility makes it cannot follow the movement of electrons and then an S-shape curve is exhibited. Therefore, the essence of hysteresis is caused by ions movement in the solar cells. The illumination plays the following roles: accelerating ions mobility; speeding up the decomposition of perovskite materials, and then leading to more ions; establishing a photo-generated electric field; increasing number of carriers in solar cells.

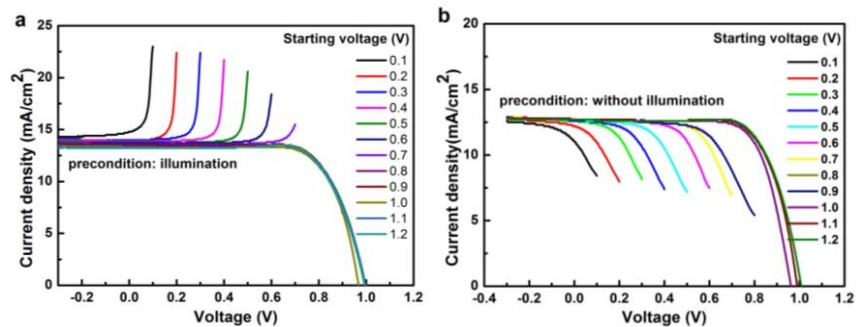

Fig.5 Influence of pre-illumination on *J-V* curves (a) with pre-illumination; (b) without pre-illumination. The starting voltage varies from 0.1 to 1.2 V.

From above results, the external applied voltage bias exhibits considerable impact on the ions migration independent of the light condition. Reverse voltage bias strengthens the ions induced electric field and accelerates ions movement, which is detrimental for the stability of solar cells. Conversely, the forward bias counteracts the ions induced electric field and delays the ion migration, which is favorable to extend the lifetime of perovskite solar cells. To confirm this hypothesis, the conversion efficiency degradation under forward (+1V) and reverse (-1V) voltage bias is tested. Figure 6 shows the normalized solar cell parameters degradation under forward and reverse bias over illumination time.

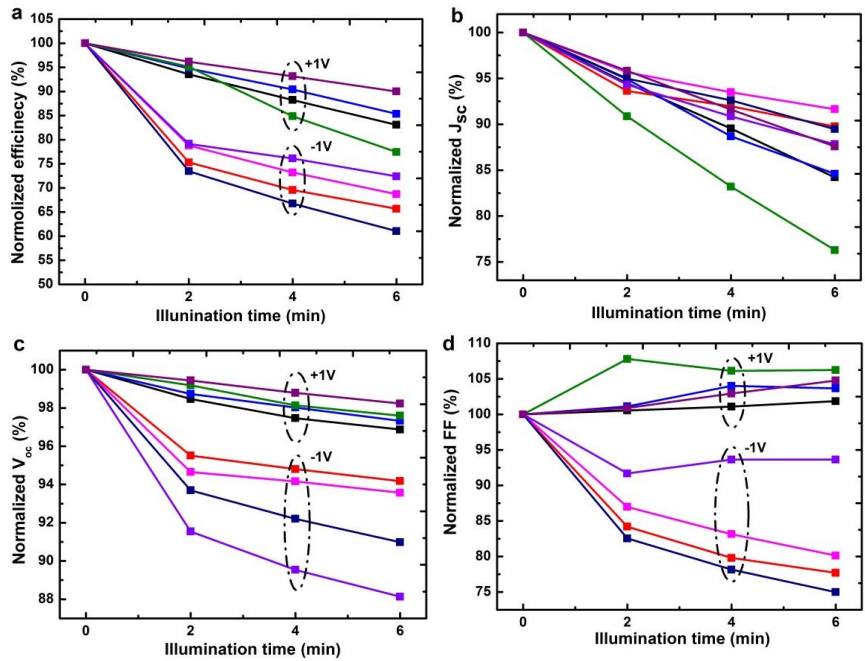

Fig.6 Conversion efficiency degradation over time under forward bias and reverse bias. Detailed normalized parameters of (a) Conversion efficiency;(b) $J_{sc}$; (c) $V_{oc}$; (d) Fill factor (FF).

Evidently, conversion efficiencies for cells under reverse bias show faster degradation than that of cells under forward bias, as shown in Fig. 6a. From Fig. 6b, the $J_{sc}$ degradation shows identical trend under reverse and forward bias. However, the degradation trend of $V_{oc}$ (Fig. 6c) exhibits almost the same with that of conversion efficiency does, indicating that the applied voltage bias mainly causes the alteration of ions induced electric field and then the conversion efficiency. The intriguing result is exhibited in fill factor (*FF*) degradation, as shown in Fig. 6d, i.e. *FF* presents enhancement under forward bias while it shows reduction under reverse bias. The possible reason is that the forward bias compensates the ions induced electric field, and then leads to enhancement of total electric field in device. Hence, the carrier collection efficiency raises up. In contrast, reverse bias results in promotion of ions induced electric field, causing reduction of total electric field. Thus, *FF* exhibits significant decrease. These findings also give us a revelation for how to extend the lifetime of perovskite solar cells. Fortunately, solar cell works on the maximum power output point (under forward bias), so it is feasible to realize stable perovskite solar cells.

**The capacitance evolution caused by ions migration.**

The last question is: Does the ions migration influence the capacitance. To investigate the capacitance evolution, the measurement of *C-V* is conducted at dark condition. Taking into account the slow mobility of ions, AC signal frequency is changed from 100 Hz to several thousand Hz. Figure 7a shows the *C-V* characteristics for the solar cell shown in Fig. 2a at different frequencies. It is found that crooks in the *C-V* curves are observed when the frequency is lower than 800 Hz. Especially, in the case of 100 and 200 Hz, the capacitance converts from positive to negative when the voltage over the inherent electric field.

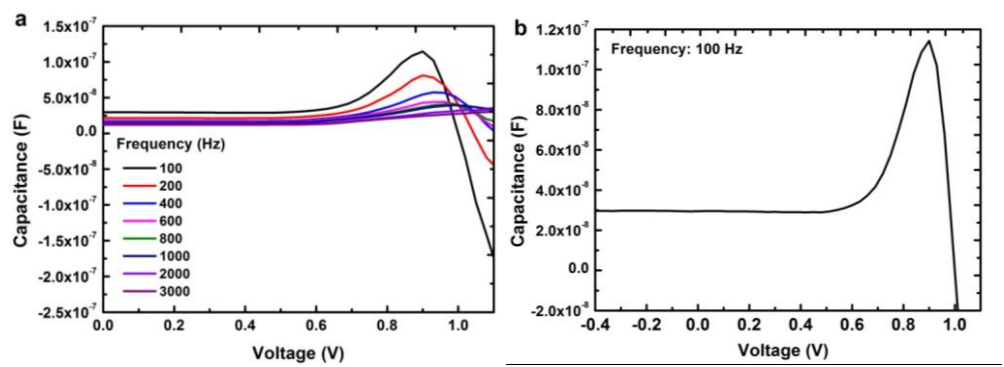

Fig.7 (a) C-V characteristics under various frequency.

(b) C-V characteristic under frequency of 100 Hz.

Asif Islam Khan et al. experimentally demonstrated the negative capacitance in a ferroelectric film, which was attributed to polarization reversal induced by electric field[37, 38]. From Fig.2d, we also find that $MA^+$ and $I^-$ ions exchange positions compared with their initial state, which is possibly responsible for the conversion of capacitance from positive to negative. For a clear view, *C-V* curve under 100 Hz is drawn in Fig. 7b. The capacitance reduces when the applied bias sweeps from minus to plus and arrives at minimum when the voltage is around 0.5 V. This result indicates that the capacitance changes at 0.5 V, consistent with the forward bias value at which the EQE is largest. Both results prove that the ion-induced electric field is about 0.5 V. Obviously, ions make significant contribution to the capacitance at low frequency. This is also the possible reason of giant dielectric response at low frequency. However, in the case of high frequency, the crook disappears because ions cannot respond to the applied measurement signal. The electrons and holes dominate the contribution to the charge.

**Discussion**

Briefly, an additional ions induced electric field is formed in the perovskite solar cells owing to the ions migration under the built-in potential. Both the built-in and ions induced electric fields govern the movement of ions and electrons. When an external voltage bias is applied on the device, the electric fields rearrange, causing movement of ions and electrons. In the dark *J-V* measurement, $MA^+$ and $I^-$ will move back to their initial locations as soon as the forward voltage bias applied on the device. Simultaneously, electrons will move and contribute to the detected current. Owing to the inconsistent mobility of ions and electrons, a crook is detected on *J-V* curves when applied voltage delay time is shorter than time duration of ion migration (around 12s). In the photo *J-V* curve, the crook is disappeared because illumination accelerates the migration of ions or the photo-generated field eliminates the ions induced field. Meanwhile, forward bias eliminates the ion-induced field while reverse filed strengthens it, which results in the cell performance recorded at reverse scan direction higher than that obtained at forward scan direction. Another factor to affect the hysteresis is the cell structure. For mesoporous structure, the device involves two perovskite layers: a perovskite infiltrated mesoporous layer and a capped perovskite layer. The infiltrated perovskite layer is difficult to be induced by the built-in electric field, hysteresis is mainly originated from capped perovskite layer. The thicker the mesoporous layer is, the slighter of hysteresis is. For the planar structured solar cell, only the perovskite layer is sandwiched by p-type and n-type layer, where the whole layer will be induced. This is why the solar cell with planar structure presented more serious hysteresis. The most critical factor to improve the stability of perovskite solar cells is to fabricate high quality stoichiometric perovskite layer. The excess I⁻ and/or MA⁺ will migrate under built-in potential, resulting in an induced electric field. The ions movement is the origin of unusual hysteresis. Optimal fabrication process and cell structure remain need to explore. At last, the novel negative capacitance observed in perovskite material will further broaden its application in optoelectronic devices.

In summary, our study initially demonstrates the ions induced electric field in perovskite solar cells. The hysteresis is preferably attributed to the migration of ions. The applied voltage exhibited considerable influence on stability of perovskite solar cells. Meanwhile, the negative capacitance is firstly detected under forward bias.

**Methods**

**Materials preparation**. MAI was synthesized by reacting methylamine with hydroiodic acid according to the literatures[39, 40]. Briefly, adding 30 mL methylamine (40% in methanol, Tokyo Chemical Industry) into 28.4 mL hydroiodic acid (57% in water, Tokyo Chemical Industry) by dropwise, the hydroiodic acid was kept in ice water (0℃) during whole process. After keeping stirring for 2 hours, the methanol and water in reactant solution were evaporated at 60℃. The precipitate was washed using diethyl ether several times. Recrystallization was performed by mixed solutions of ethanol and diethyl ether. The precipitated crystal of MAI was dried before use. FAI was prepared by adding 3 g formamidine acetate (Tokyo Chemical Industry) into 13 mL HI (57 wt% in water), the reaction was finished at 50℃[41]. The washing and recrystallization processes were same as MAI.

**Perovskite solar cell fabrication**. Mesoposous structured perovskite solar cells were prepared for investigations. First, a series of cleanings of deionized water, detergent, acetone, methanol and UV-ozone were sequentially conducted. Second, around 50 nm blocking $TiO_2$ layer was deposited on FTO substrates (TEC7, Pilkington) by spray pyrolysis at 550℃ using mixed solvents of 0.3 mL Titanium diisopropoxide bis(acetylacetonate) (Sigma-Aldrich) and 4 mL ethanol. Third, mesoporous $TiO_2$ layer was prepared by spin coating with $TiO_2$ nanoparticles (PST-18NR) based solution. Before sintering treatment at 550℃ for 15 min, the sample was dried at 125℃ about 5 min. Fourth, PbI$_2$ powder was dissolved in DMF and DMSO mixed solvents with volume ratio of 9:1 and the solution concentration was 553 mg/mL. PbI$_2$ film was prepared by spin coating at rotation speed of 3000 rpm 5 s and 6000 rpm 5s. After annealing treatments at 70℃ for 5 min and 100 ℃ for 5 min, 200 uL MAI ( or FAI+MAI) solution with concentration of 6 mg/mL was dropped on PbI$_2$ films to form perovskite layer. The loading time of 2 min was required before spin coating at 4000 rpm for 25 s. Annealing at 100 ℃ for 5 min was conducted. Then, Spiro-OMeTAD solution with concentration of 72.25 mg/ml in chlorobenzene with additive of 17.5 ul lithium bis(trifluoro methanesulfonyl)imide (520 mg/mL in acetonitrile, Sigma-Aldrich) and 26.25 uL 4-tert-butylpyridine (Sigma-Aldrich) was spin coated and followed 15 min annealing at 70 ℃. At last, Au electrode was deposited by thermal evaporation and the active area was 0.12 cm$^2$.

**Solar cell Characterization**. The *J-V* curves at dark and under illumination (AM1.5, 100 mW/cm$^2$) were recorded at the required scan rate and delay time. *EQE* spectra were measured using an *EQE* system (CEP-25RR) under AC mode with chopping frequency of 15 Hz at various voltage biases. The excitation light intensity is calibrated at 50 μW/cm$^2$ by a Si solar cell. For the conversion efficiency degradation measurement, the solar cells are irradiated under one sun with forward or reverse bias. After required time, the *J-V* curves are recorded. The *C-V* characteristic was measured with a small voltage perturbation of 10 mV at dark on a Precision LCR Meter (Hewlett Packard 4284A).

**Reference**

1. Kojima, A., Teshima, K., Shirai, Y. & Miyasaka, T. Organometal Halide Perovskites as Visible-Light Sensitizers for Photovoltaic Cells, *J. Am. Chem. Soc.* **131**, 6050-6051 (2009).
2. Yang, W. S., Noh, J. H., Jeon, N. J., Kim, Y. C., Ryu, S., J. Seo, Seok, S. Il high-performance photovoltaic perovskite layers fabricated through intramolecular exchange, *Science* **348**, 1234-1237 (2015).
3. Jackson, P., Hariskos, D., Wuerz, R., Kiowski, O., Bauer, A., Friedlmeier, T. M., & Powalla, M., Properties of Cu(In,Ga)Se$_2$ solar cells with new record efficiencies up to 21.7%, *Phys. Status Solidi RRL* **9**, 28–31 (2015).

4. PV Magazine. First Solar raises bar for CdTe with 21.5% efficiency record, 6 February 2015.

5. Leijtens, T. G., Eperon, E., Pathak, S., Abate, A., Lee, M. M., & Snaith, H. J., Overcoming ultraviolet light instability of sensitized $TiO_2$ with meso-superstructured organometal tri-halide perovskite solar cells, *Nature Communications*, **4**, 2885 (2013).

6. Habisreutinger, S. N., Leijtens, T., Eperon, G. E., Stranks, S. D., Nicholas, R. J., & Snaith, H. J. Carbon Nanotube/Polymer Composites as a Highly Stable Hole Collection Layer in Perovskite Solar Cells, *Nano Lett.* **14**, 5561–5568 (2014).

7. Mei, A., Li, X., Liu, L., Ku, Z., Liu, T., Rong, Y., Xu, M., Hu, M., Chen, J., Yang, Y., Grätzel, M., & Han, H. A hole-conductor–free, fully printable mesoscopic perovskite solar cell with high stability, *Science* **345**, 295-296 (2014).

8. Snaith, H. J., Abate, A., Ball, J. M., Eperon, G. E., Leijtens, T., Noel, N. K., Stranks, S. D., Wang, J. Tse-Wei, Wojciechowski, K., Zhang, W. Anomalous Hysteresis in Perovskite Solar Cells, *J. Phys. Chem. Lett.* **5**, 1511–1515 (2014).

9. Tress, W., Marinova, N., Moehl, T., Zakeeruddin, S. M., Nazeeruddin, M. K. & Grätzel, M. Understanding the rate-dependent J–V hysteresis, slow time component, and aging in $CH_3NH_3PbI_3$ perovskite solar cells: the role of a compensated electric field, *Energy Environ. Sci.*, **8**, 995–1004 (2015).

10. Zhang, Y., Liu, M., Eperon, G. E., Leijtens, T. C., McMeekin, D., Saliba, M., Zhang, W., Bastiani, M. de, Petrozza, A., Herz, L. M., Johnston, M. B., Lin, H. & Snaith, H. J. Charge selective contacts, mobile ions and anomalous hysteresis in organic–inorganic perovskite solar cells, *Materials Horizons*, **2**, 315-322 (2015).

11. Unger, E. L., Hoke, E. T., Bailie, C. D., Nguyen, W. H., Bowring, A. R., Heumuller, T., Christoforod, M. G. & McGehee, M. D. Hysteresis and transient behavior in current–voltage measurements of hybrid-perovskite absorber solar cells, *Energy Environ. Sci.*, **7**, 3690-3698 (2014).

12. Frost, Jarvist M., Butler, Keith T. & Walsh, A. Molecular ferroelectric contributions to anomalous hysteresis in hybrid perovskite solar cells, *APL Materials* **2**, 081506 (2014).

13. Shao, Y., Xiao, Z., Bi, C., Yuan, Y. & Huang, J. Origin and elimination of photocurrent hysteresis by fullerene passivation in $CH_3NH_3PbI_3$ planar heterojunction solar cells, *Nature Communications*, **5**, 5784 (2014).

14. Kim, H. & Park, N. Parameters Affecting I–V Hysteresis of $CH_3NH_3PbI_3$ Perovskite Solar Cells: Effects of Perovskite Crystal Size and Mesoporous $TiO_2$ Layer, *J. Phys. Chem. Lett.* **5**, 2927–2934 (2014).

15. Zhao, C., Chen, B., Qiao, X., Luan, L., Lu, K. & Hu, B. Revealing Underlying Processes Involved in Light Soaking Effects and Hysteresis Phenomena in Perovskite Solar Cells, *Adv. Energy Mater.* **5**, 1500279 (2015).

16. Ono, L. K., Raga, S. R., Wang, S., Kato, Y. & Qi, Y. Temperature-dependent hysteresis effects in perovskite-based solar cells, *J. Mater. Chem. A,* **3**, 9074–9080 (2015).

17. Zhang, H., Liang, C., Zhao, Y., Sun, M., Liu, H., Liang, J., Li, D., Zhang, F. & He, Z. Dynamic interface charge governing the current–voltage hysteresis in perovskite solar cells, *Phys. Chem. Chem. Phys.*, **17**, 9613-9618 (2015).

18. Jena, A. K., Chen, H., Kogo, A., Sanehira, Y., Ikegami, M. & T. Miyasaka, The Interface between FTO and the $TiO_2$ Compact Layer Can Be One of the Origins to Hysteresis in Planar Heterojunction Perovskite Solar Cells, *ACS Appl. Mater. Interfaces* **7**, 9817–9823 (2015).

19. Ip, A. H., Quan, L. N., Adachi, M. M., McDowell, J. J., Xu, J., Kim, D. H. & Sargent, E. H. A two-step route to planar perovskite cells exhibiting reduced hysteresis, *Appl. Phys. Lett.* **106**, 143902 (2015).

20. Liu, C., Fan, J., Zhang, X., Shen, Y., Yang, L. & Mai, Y. Hysteretic Behavior upon Light Soaking in Perovskite Solar Cells Prepared via Modified Vapor-Assisted Solution Process, *ACS Appl. Mater. Interfaces* **7**, 9066–9071 (2015).

21. Xiao, Z., Yuan, Y., Shao, Y., Wang, Q., Dong, Q., Bi, C., Sharma, P., Gruverman, A. & Huang, J. Giant switchable photovoltaic effect in organometal trihalide perovskite devices, *Nature Materials*, **14**, 193–198 (2015).

22. Yuan, Y., Chae, J., Shao, Y., Wang, Q., Xiao, Z., Centrone, A. & Huang, J. Photovoltaic Switching Mechanism in Lateral Structure Hybrid Perovskite Solar Cells, *Adv. Energy Mater.* **5**, 1500615, 2015.

23. Eames, C., Frost, J. M., Barnes, P. R. F., O'Regan, B. C., Walsh, A. & Islam, M. S. Ionic transport in hybrid lead iodide perovskite solar cells, *Nature Communications*, **6**, 7497 (2015).

24. O'Regan, B. C., Barnes, P. R. F., Li, Xiaoe, Law, Chunhung, Palomares, Emilio &

Marin-Beloqui, Jose M. Optoelectronic Studies of Methylammonium Lead Iodide Perovskite Solar Cells with Mesoporous $TiO_2$: Separation of Electronic and Chemical Charge Storage, Understanding Two Recombination Lifetimes, and the Evolution of Band Offsets during J–V Hysteresis, *J. Am. Chem. Soc.* **137**, 5087–5099 (2015).

25. Salim, T., Sun, S., Abe, Y., Krishna, A., Grimsdalea, A. C. & Lam, Y. M. Perovskite-based solar cells: impact of morphology and device architecture on device performance, *J. Mater. Chem. A*, **3**, 8943-8969 (2015).

26. Mizusaki, J., Arai, K. & Fueki, K. Ionic conduction of the perovskite-type halides, *Solid State Ionics*, **11**, 203–211 (1983).

27. Alberecbt, M. G. & Green, M. The kinetics of the photolysis of thin films of lead iodide, *J. Phys. Chem. Solids.* **38**, 297-306 (1977).

28. Yamada, K., Isobe, K., Tsuyama, E., Okuda, T. & Furukawa, Y., Ionic conduction of the perovskite-type halides, *Solid State Ionics*, **79**, 152–157 (1995).

29. Verwey, J. F. Time and intensity dependence of the photolysis of lead halides, *J. Phys. Chem. Solids,* **31**, 163–168 (1970).

30. Chen, Q., Mao, L., Li, Y., Kong, T., Wu, N., Ma, C., Bai, S., Jin, Y., Wu, D., Lu, W., Wang, B. & Chen, L. Quantitative operando visualization of the energy band depth profile in solar cells, Nature Communications, **6**, 7745 (2015).

31. Zhou, H., Chen, Q., Li, G., Luo, S., Song, T., Duan, H., Hong, Z., You, J., Liu, Y. & Yang, Y. Interface engineering of highly efficient perovskite solar cells, *Science*, **345**, 542-546 (2014).

32. Wojciechowski, K., Leijtens, T., Spirova, S., Schlueter, C., Hoerantner, M., Tse-Wei Wang, J., Li, C., Jen, A. K.-Y., Lee, T. & Snaith, H. J. C60 as an Efficient N-Type Compact Layer in Perovskite Solar Cells, *J. Phys. Chem. Lett.*, **6**, 2399–2405 (2015).

33. Chen, H., Sakai, N., Ikegami, M. & Miyasaka, T. Emergence of Hysteresis and Transient Ferroelectric Response in Organo-Lead Halide Perovskite Solar Cells, *J. Phys. Chem. Lett.* **6**, 164–169 (2015).

34. Sanchez, R. S., Gonzalez-Pedro, V., Lee, J., Park, N., Kang, Y., Mora-Sero, I. & Bisquert, J. Slow Dynamic Processes in Lead Halide Perovskite Solar Cells. Characteristic Times and Hysteresis, *J. Phys. Chem. Lett.* **5**, 2357–2363 (2014).

35. Wei, J., Zhao, Y., Li, H., Li, G., Pan, J., Xu, D., Zhao, Q. & Yu, D. Hysteresis Analysis

Based on the Ferroelectric Effect in Hybrid Perovskite Solar Cells, *J. Phys. Chem. Lett.* **5**, 3937–3945 (2014).

36. Hoke, E. T., Slotcavage, D. J., Dohner, E. R., Bowring, A. R., Karunadasea, H. I., McGehee, M. D. Reversible photo-induced trap formation in mixed-halide hybrid perovskites for photovoltaics, *Chem. Sci.*, **5**, 613-617 (2015).
37. Khan, A. I., Chatterjee, K., Wang, B., Drapcho, S., You, L., Serrao, C., Bakaul, S. R., Ramesh, R. & Salahuddin, S. Negative capacitance in a ferroelectric capacitor, *Nature materials*, **14**, 182-186 (2015).
38. Catalan, G., Jiménez, D. & Gruverman, A. Negative capacitance detected, *Nature materials*, **14**, 137-139 (2015).
39. Etgar, L., Gao, P., Xue, Z., Peng, Q., Chandiran, A. K., Liu, B., Nazeeruddin, Md. K. & Grätzel, M. Mesoscopic $CH_3NH_3PbI_3/TiO_2$ Heterojunction Solar Cells, *J. Am. Chem. Soc.* **134**, 17396-17399 (2012).
40. Ito, S., Tanaka, S., Vahlman, H., Nishino, H., Manabe, K. & Lund, P. Carbon-Double-Bond-Free Printed Solar Cells from $TiO_2/CH_3NH_3PbI_3/$ CuSCN/Au: Structural Control and Photoaging Effects, *Chem. Phys. Chem* **15**, 1194 – 1200 (2014).
41. Hu, M., Liu, L., Mei, A., Yang, Y., Liu, T. & Han, H. Efficient hole-conductor-free, fully printable mesoscopic perovskite solar cells with a broad light harvester $NH_2CH=NH_2PbI_3$, *J. Mater. Chem. A*, **2**, 17115-17121 (2014).

**Acknowledgements**

The authors would like to thank Dr. Jakapan Chantana in Ritsumeikan University, Dr. Qiang Chen in Xiamen University, Dr. Yi Ding in Nankai University and Dr. Qiming Liu in Saitama University for their useful discussions.

# Electronic Supplementary Information

# Experimental demonstration of ions induced electric field in perovskite solar cells

Zeguo Tang[1,*] & Takashi Minemoto [2]

[1]Ritsumeikan Global Innovation Research Organization, Ritsumeikan University,

1-1-1 Nojihigashi, Kusatsu, Shiga 525-8577, Japan

[2]Department of Electrical and Electronic Engineering, Ritsumeikan University,

1-1-1 Nojihigashi, Kusatsu, Shiga 525-8577, Japan

Email: tangzg@fc.ritsumei.ac.jp

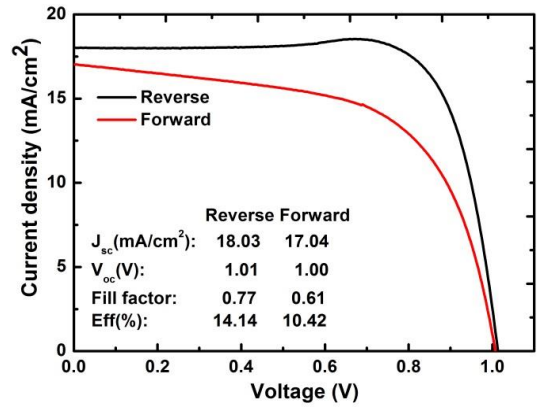

**Supplementary Figure 1.** *J-V* curve for perovskite solar cell for measuring EQE spectrum shown in Fig. 2. The delay time is fixed at 30 ms while record this curve.

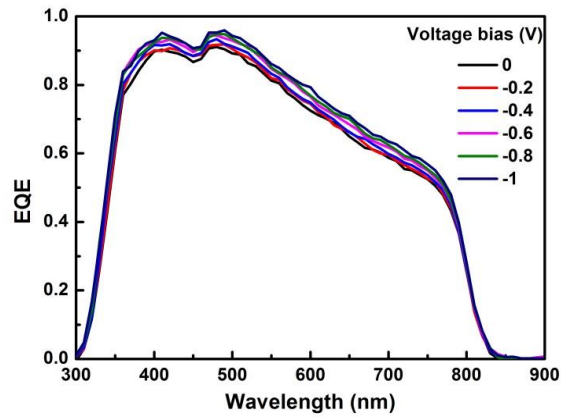

**Supplementary Figure** 2. EQE spectra under reverse bias for perovskite solar cell.

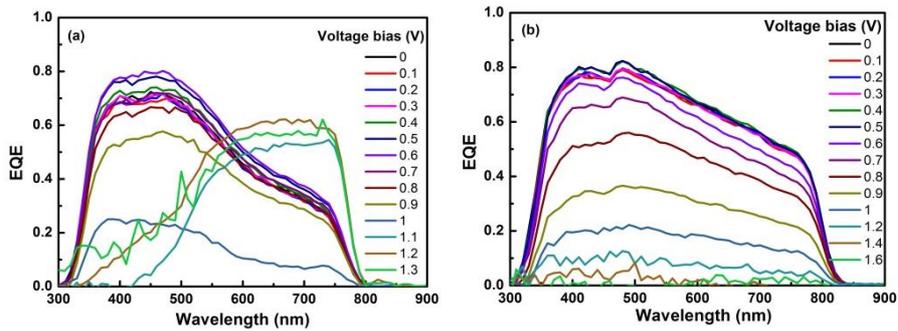

**Supplementary Figure 3**. EQE spectra under forward bias for perovskite solar cells based on (a) $MAPbI_3$ and $(FAPbI_3)_{0.85}$ $(MAPbI_3)_{0.15}$ materials.

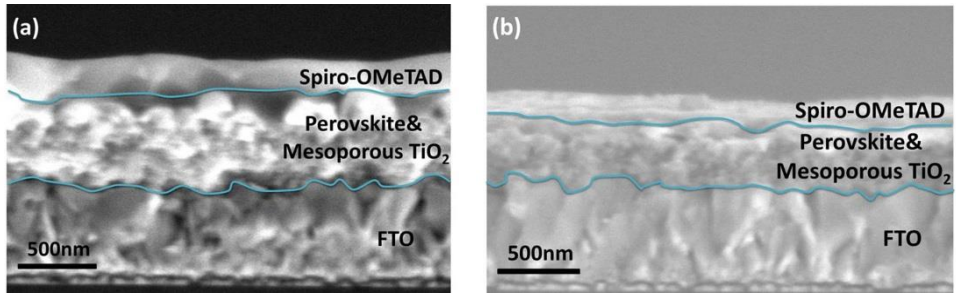

**Supplementary Figure** 4. Cross-sectional SEM images for perovskite solar cells with unequal thickness perovskite layers.

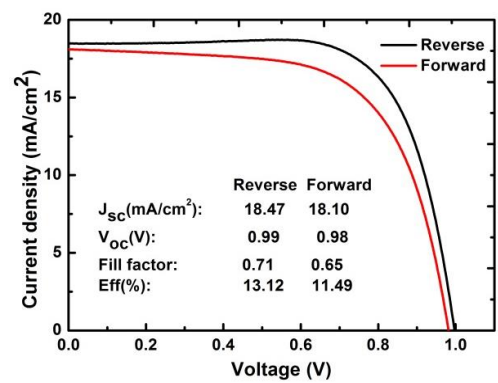

**Supplementary Figure 5**. *J-V* curve for perovskite solar cell with thin perovskite layer. . The delay time is fixed at 30 ms while record this curve.